\documentclass[aps,prd,twocolumn]{revtex4}
\usepackage{graphicx}

\begin{document}

\title{The Emergent Universe:\\
Inflationary cosmology with no singularity}

\author{George F.R. Ellis$^1$ and Roy Maartens$^{1,2}$}

\affiliation{~}

\affiliation{$^1$Department of Mathematics \& Applied Mathematics,
University of Cape Town, Cape~Town~7701, South~Africa}

\affiliation{$^2$Institute of Cosmology \& Gravitation, University
of Portsmouth, Portsmouth~PO1~2EG, UK}

\begin{abstract}
Observations indicate that the universe is effectively flat, but
they do not rule out a closed universe. The role of positive
curvature is negligible at late times, but can be crucial in the
early universe. In particular, positive curvature allows for
cosmologies that originate as Einstein static universes, and then
inflate and later reheat to a hot big bang era. These cosmologies
have no singularity, no ``beginning of time", and no horizon
problem. If the initial radius is chosen to be above the Planck
scale, then they also have no quantum gravity era, and are
described by classical general relativity throughout their
history.
\end{abstract}

\maketitle

\section{Introduction}

The ``standard" inflationary model is based on a flat ($ K=0
\Leftrightarrow \Omega _{0}=1$) Friedmann-Robertson-Walker (FRW)
geometry, motivated by the fact that inflationary expansion
rapidly wipes out any original spatial curvature. However, even
though inflation drives the curvature term,
\begin{equation}
\Omega (t)-1={\frac{K}{a^{2}H^{2}}}\,
\end{equation}
towards zero, or equivalently, drives the total density parameter
$\Omega $ towards 1, this does \emph{not} imply that $K=0$.
Conditions leading to the open set of values $\Omega _{0}>1$ are
far less fine-tuned than those corresponding precisely to $K=0$
(although closed models have other fine-tuning aspects). But
irrespective of any arguments about fine-tuning, the spatial
curvature of the real universe is in principle determined by
observations; theory will have to give way to data if the data
clearly tell us that $\ \Omega _{0}>1$. Recent cosmic microwave
background (CMB) and other data~\cite{cmb} are not conclusive, but
include the possibility that $K=+1$, with
\begin{equation}
\Omega _{0}=1.02\pm 0.02\,\,.
\end{equation}

Future experiments such as PLANCK will reduce these error bars and
give more accurate information about the curvature of the
universe.
(Note that the true global value of $\Omega_0$ and its observed
value in the local Hubble volume may differ due to cosmological
perturbations seeded by inflation.) Current data mean that we
should take closed universes, and therefore closed inflationary
models, seriously. Inflation in a closed universe is sometimes
considered to be ruled out by unrealistic
fine-tuning~\cite{linde}. If this were true, then observational
confirmation of $\Omega_0>1$ would rule out inflation. But in fact
there is a wide range of closed inflationary models, and it is not
too difficult to construct simple and consistent models without
excessive fine-tuning (see also Ref.~\cite{ld}).
It is also true that fine-tuned initial conditions cannot be ruled
out by purely scientific arguments, as we discuss below. For the
models that we consider, the value of $\ \Omega _{0}-1$ is set by
choosing one parameter, while the magnitude of large-scale scalar
perturbations fixes another parameter.

If $\Omega _{0}$ is taken as 1.02, then the power spectra of CMB
anisotropies and matter can show testable differences from the
standard flat model~\cite{ld}. If $\ \Omega _{0}-1$ is very small,
then curvature is negligible as regards structure formation
processes in the universe. The smaller that $\Omega_{0}-1$ is, the
more closely the models approximate the standard flat model. But
in the early inflationary universe, positive curvature can play a
significant role and lead to novel features that do not arise when
$K\leq 0$, for example allowing a minimum in the scale factor
$a(t)$ which is otherwise not possible, or putting limits on the
number of possible e-foldings in the inflationary
epoch~\cite{infplus1,uke}. Primordial history is very different
when $K=+1$ than in the case $\Omega _{0}=1$, regardless of how
small $\Omega _{0}-1$ is, and a variety of possibilities arise,
including universes with a minimum radius.

Singularity theorems have been devised that apply in the
inflationary context, showing that the universe necessarily had a
beginning (according to classical and semi-classical
theory)~\cite{bv1}. In other words, according to these theorems
the quantum gravity era cannot be avoided in the past even if
inflation takes place. However, the models we present escape this
conclusion, because they do not satisfy the geometrical
assumptions of these theorems. Specifically, the theorems assume
either (a)~that the universe has open space sections, implying
$K=0$ or $-1$, or (b)~that the Hubble expansion rate $H=\dot{a}/a$
is bounded away from zero in the past, $H>0$. There are
inflationary universes that evade these constraints and hence
avoid the conclusions of the theorems (this was also noted in
Ref.~\cite{mv} ). It is also possible to find counter-examples
that are open, i.e., where assumption (a) is not violated, but
this typically involves sophisticated constructions~\cite{ag}.

Here we consider closed models in which $K=+1$ and $H$ can become
zero, so that both assumptions (a) and (b) of the inflationary
singularity theorems are violated. The models are simple, obey
general relativity, and contain only ordinary matter and
(minimally coupled) scalar fields. Previous examples of closed
inflationary models~\cite{infplus1,uke,mv,bounce} are, to our
knowledge, either bouncing models or models in which inflation is
preceded by deceleration (so that a singularity is not avoided).
The $K=+1$ bouncing universe collapses from infinite size in the
infinite past and then turns around at $t_{{\rm i} }$ to expand in
an inflationary phase. The canonical model for such a bounce is
the de~Sitter universe in the $K=+1$ frame, with $ a(t)=a_{{\rm i}
}\cosh Ht\,.$ These coordinates cover the whole spacetime, which
is geodesically complete~\cite{sch}. However, the bouncing models
face serious difficulties as realistic cosmologies. The initial
state is hard to motivate (collapsing from infinite size without
causal interaction), and it is also difficult to avoid
nonlinearities in the collapse that prevent a regular bounce.

\section{The Emergent Universe scenario}

We show here that when $\ K=+1$ there are closed inflationary
models that do not bounce, but inflate from a static beginning,
and then reheat in the usual way. The inflationary universe
emerges from a small static state that has within it the seeds for
the development of the macroscopic universe, and we call this the
``Emergent Universe" scenario. (This can be seen as a modern
version and extension of the Eddington universe.) The universe has
a finite initial size, with a finite amount of inflation occurring
over an infinite time in the past, and with inflation then coming
to an end via reheating in the standard way. The redshift and the
total number of e-folds remain bounded through the expansion of
the universe until the present day, because the scale-factor is
bounded away from zero in the past. There is {\em no} horizon
problem, since the initial state is Einstein static. Since they
start as Einstein static, they avoid a singularity. The initial
static state can be chosen to have a radius above the Planck
scale, so that these models can even avoid a quantum gravity
regime, whatever the true quantum gravity theory may be.

Because they can undergo a large amount of inflation, these models
can be effectively the same as the standard flat models as regards
structure formation processes. Therefore they are not vulnerable
to future reductions in the observed value of $\Omega_0-1$ below
0.02 (provided that it remains positive), and they can reproduce
the successes of the standard inflationary cosmologies, but from
very different primordial foundations.

We do not require exotic physics or matter. The early universe
contains a standard scalar field $\phi $ with energy density $\rho
_{\phi }= \frac{1}{2}\dot{\phi}^{2}+V(\phi )$ and pressure
$p_{\phi }= \frac{1}{2}\dot{\phi}^{2}-V(\phi )$, and possibly also
ordinary matter with energy density $\rho $ and pressure $p=w\rho
$, where $-{\frac{1 }{3}}\leq w\leq 1$. The cosmological constant
is absorbed into the potential $V$. There are no interactions
between matter and the scalar field, so that they separately obey
the energy conservation and Klein-Gordon equations,
\begin{eqnarray}
\dot{\rho}+3(1+w)H\rho &=&0\,,  \label{cons} \\
\ddot{\phi}+3H\dot{\phi}+V^{\prime }(\phi ) &=&0\,.  \label{kg}
\end{eqnarray}
The Raychaudhuri field equation
\begin{equation}
{\frac{\ddot{a}}{a}}=-{\frac{8\pi G}{3}}\left[
{\frac{1}{2}}(1+3w)\rho +\dot{ \phi}^{2}-V(\phi )\right] ,
\label{Ray_Fr}
\end{equation}
has first integral the Friedmann equation,
\begin{equation}
H^{2}={\frac{8\pi G}{3}}\left[ \rho
+{\frac{1}{2}}\dot{\phi}^{2}+V(\phi ) \right]
-{\frac{K}{a^{2}}}\,,  \label{Friedmann}
\end{equation}
which together imply
\begin{equation}
\dot{H}=-4\pi G\left[ \dot{\phi}^{2}+(1+w)\rho \right]
+{\frac{K}{a^{2}}}\,. \label{b1}
\end{equation}
The Raychaudhuri equation gives the condition for inflation,
\begin{equation}
{\ddot{a}}>0~\Leftrightarrow
~\dot{\phi}^{2}+{\frac{1}{2}}(1+3w)\rho <V(\phi )\,.  \label{2}
\end{equation}

For a positive minimum in the inflationary scale factor, $a_{\rm
i} \equiv a(t_{\rm i})>0$,
\begin{equation}  \label{3}
H_{{\rm i}}=0 ~\Leftrightarrow~ {\frac{1}{2}}\dot{\phi}_{\rm i}^2
+V_{\rm i}+\rho_{\rm i} = { \frac{3K }{8\pi Ga^2_{{\rm i}}}}\,,
\end{equation}
where the time $t_{{\rm i}}$ may be infinite. The only way to
satisfy Eq.~(\ref{3} ) with non-negative energy densities is if
$K=+1$. Closed inflationary models admit a minimum scale factor if
inflation occurs for long enough, since curvature will eventually
win over a slow-rolling scalar field as we go back into the past
(cf.~\cite{infplus1}). The inflationary singularity theorems
mentioned above exclude this case, since they either only consider
$ K\leq0$, or explicitly exclude the possibility $H_{{\rm i}}=0$.

Closed models with a minimum scale factor $a_{{\rm i}}>0$ include
both bouncing and ever-expanding cases. We do not proceed further
with the bouncing models because of their acausal initial state
and the difficulty of achieving a stable bounce. Ever-expanding
models avoid these problems. A simple ever-inflating model is the
closed model containing radiation ($w={\frac{1}{3}}$) and
cosmological constant $\Lambda= 8 \pi GV$, with scale
factor~\cite{h}
\begin{equation}  \label{rad}
a(t)=a_{{\rm i}}\left[1+ \exp\left({\frac{\sqrt{2}\,t }{a_{{\rm
i}}}} \right) \right] ^{1/2}.
\end{equation}
In the infinite past, $t\to-\infty$, the model is asymptotically
Einstein static, $a\to a_{\rm i}$. Inflation occurs for an
infinite time to the past, but at any finite time
$t_{\mathrm{e}}$, there is a finite number of e-folds,
\begin{equation}  \label{rad2}
N_{\mathrm{e}}=\ln {\frac{a_{\mathrm{e}} }{a_{\rm i}}}\approx
{\frac{t_{\mathrm{e} } }{\sqrt{2} a_{\rm i}}}\,,
\end{equation}
where the last equality holds for $t_{\mathrm{e}}\gg a_{\rm i}$.
The curvature parameter at $t_{\mathrm{e}}$ is strongly suppressed
by the de Sitter-like expansion:
\begin{equation}
\Omega_{\mathrm{e}}-1\approx 2e^{-N_{\mathrm{e}}}\,.
\end{equation}

The exact model Eq.~(\ref{rad}) is a simple example of
Eddington-type solutions. There are trajectories of this type in
the classical phase space~\cite{em} satisfying Einstein's
equations. They are past-asymptotically Einstein static and
ever-expanding, and the fact that they exist already shows there
are inflationary universes that evade the conditions of the
singularity theorems presented in~\cite{bv1}. However, in these
models inflation does not end. Below we discuss more realistic
models, based on inflationary potentials, that do exit from
inflation. These universes are singularity-free, without particle
horizons, and ever-expanding ($H\geq 0$). Even though Emergent
Universes admit closed trapped surfaces, these do not lead to a
singularity, since $K=+1$ and the weak energy condition is
violated in the past~\cite{ects}.

The Einstein static universe is characterized by $K=+1$ and
$a=a_{\mathrm{i} }=\,$const. Equations~(\ref{cons})--(\ref{b1})
then imply that
\begin{eqnarray}
{\frac{1}{2}}(1-w_{\mathrm{i}})\rho_{\mathrm{i}}+V_{\mathrm{i}}
&=&{\frac{1}{4\pi
Ga_{\mathrm{i}}^2}}\,,  \label{es1} \\
(1+w_{\mathrm{i}})\rho_{\mathrm{i}}+\dot{\phi}^2_{\mathrm{i}} &=&
{\frac{1}{ 4\pi G a_{\mathrm{i}}^2}}\,,  \label{es2}
\end{eqnarray}
where $\dot{\rho}_{\mathrm{i}}=0=\ddot{\phi}_{\mathrm{i}}$ and
$V_{ \mathrm{i}}= \Lambda_{\mathrm{i}}/ 8\pi G$ is the primordial
vacuum energy. If the scalar field kinetic energy vanishes, i.e.
if $\dot{\phi}_{\mathrm{i}}=0$, then $
(1+w_{\mathrm{i}})\rho_{\mathrm{i}}>0$, so that there must be
matter to keep the universe static. If the static universe has
only a scalar field, i.e. if $\rho_{\mathrm{i}}=0$, then the field
must have nonzero (but constant) kinetic energy, so that it rolls
at constant speed along the flat potential. (Dynamically, this
case is equivalent to a stiff fluid, $w_{\mathrm{i}}=1$, plus
cosmological constant $\Lambda_{\mathrm{i}}$~\cite{bemt}.)

The radius $a_{\mathrm{i}}$ of the initial static universe can be
chosen to be above the Planck scale,
\begin{equation}
a_{\mathrm{i}}>M_{\mathrm{p}}^{-1}\,,
\end{equation}
by suitable choice of $V_{\mathrm{i}}$,
$\dot{\phi}_{\mathrm{i}}^2$ and $ \rho_{\mathrm{i}}$ (with all of
them~$\ll M_{\mathrm{p}}^4$). Thus these models can in principle
avoid the quantum gravity era.

A simple way to realize the scenario of the Emergent Universe is
the following.

\section{A simple Emergent potential}

Consider a potential that is asymptotically flat in the infinite
past,
\begin{equation}
V(\phi) \to
V_{\mathrm{i}}~\mbox{as}~\phi\to-\infty\,,~t\to-\infty\,,
\end{equation}
but drops towards a minimum at a finite value $\phi_{\mathrm{f}}$.
The scalar field kinetic energy density is asymptotic to the
constant Einstein static value,
\begin{equation}  \label{aes}
{\frac{1}{2}}\dot\phi^2 \to {\frac{1}{2}}V_{\mathrm{i}}=
{\frac{1}{8\pi G a_{
\mathrm{i}}^2}}~\mbox{as}~\phi\to-\infty\,,~t\to-\infty\,,
\end{equation}
where we used Eqs.~(\ref{es1}) and (\ref{es2}) with
$\rho_{\mathrm{i}}=0$. Because $\dot\phi_{\mathrm{i}}\neq 0$, no
matter is needed to achieve the initial static state. The field
rolls from the Einstein static state at $ -\infty$ and the
potential slowly drops from its Einstein static original value.
Provided that $\dot\phi^2$ decreases more rapidly than the
potential, we have $V-\dot\phi^2>0$, so that the universe
accelerates, by Eq.~(\ref{2} ). Since $\ddot\phi<0$ and
$V^{\prime}<0$, while $\dot\phi>0$, the Klein-Gordon
equation~(\ref{kg}) shows that the universe is expanding ($H>0$ ).

\begin{figure}[tbp]
\begin{center}
\includegraphics[width=6.2cm,height=8cm,angle=270]{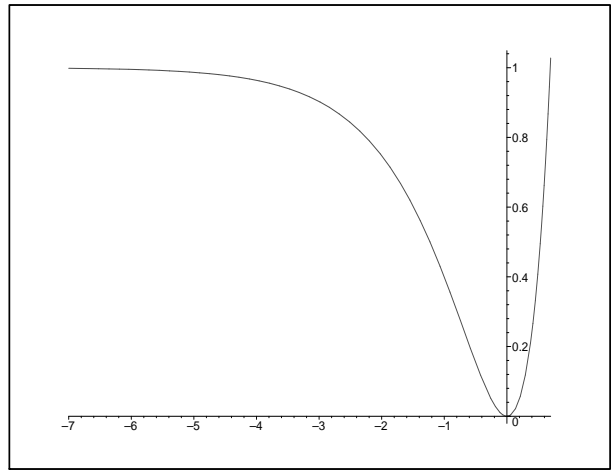}
\end{center}
\caption{ Schematic of a potential for an Emergent Universe, with
$V-V_{ \mathrm{f}}$ plotted against
$\protect\phi-\protect\phi_{\mathrm{f}}$. }
\end{figure}

Inflation ends at time $t_{\mathrm{e}}$, where
$V_{\mathrm{e}}=\dot{\phi}^2_{ \mathrm{e}}$. Then reheating takes
place as the field oscillates about the minimum at
$\phi_{\mathrm{f}}$. In the asymptotic past, $V\to V_{\mathrm{i}}$
, the primordial cosmological constant $\Lambda_{\mathrm{i}}=8\pi
GV_{ \mathrm{i}}$ is given by Eq.~(\ref{aes}) as
\begin{equation}  \label{pcc}
\Lambda_{\mathrm{i}}={\frac{2 }{a_{\mathrm{i}}^2}}\,,
\end{equation}
so that $\Lambda_{\mathrm{i}}$ is large for a small initial
radius. At the minimum, $V_{\mathrm{f}}$ defines the cosmological
constant that dominates the late universe,
\begin{equation}
\Lambda= 8\pi GV_{\mathrm{f}} ~\ll ~ \Lambda_{\mathrm{i}}\,.
\end{equation}
A typical example of a potential is shown in Fig.~1, which is
based on the potential presented in Ref.~\cite{emerge}:
\begin{equation}  \label{pot}
V-V_{\mathrm{f}}=(V_{\mathrm{i}}-V_{\mathrm{f}})\left[\exp
\left({\frac{\phi -\phi_{\mathrm{f}} }{\alpha}}\right)-1\right]^2,
\end{equation}
where $\alpha$ is a constant energy scale.

The infinite time of inflation, from $t=-\infty$ to
$t=t_{\mathrm{e}}$ ($< t_{\mathrm{f}}$), produces a finite amount
of inflation (and a finite redshift to the initial Einstein static
state). The potential produces expansion that is initially
qualitatively similar to the exact solution in Eq.~(\ref{rad}), so
that the total number of e-folds from the initial expansion can be
estimated, following Eq.~(\ref{rad2}), as
\begin{equation}
N_{\mathrm{e}}=O\left({\frac{t_{\mathrm{e}}
}{a_{\mathrm{i}}}}\right)\,.
\end{equation}
Provided that $a_{\mathrm{i}}$ is chosen small enough and
$t_{\mathrm{e}}$ large enough, a very large number of e-folds can
be produced. The parameters in the potential in Eq.~(\ref{pot})
can be chosen so that the primordial universe is consistent with
current observations~\cite{emerge}.

\section{Fine-tuning}

As with standard inflationary models, fine-tuning is necessary to
produce density perturbations at the $O(10^{-5})$ level, and to
fix $\Lambda $ so that $\Omega _{\Lambda 0}\sim 0.7$. These issues
are faced by all inflationary universe models~\cite{b}. The
specific geometrical fine-tuning problem in the Emergent models is
the requirement of a particular choice of the initial radius
$a_{\mathrm{i}}$, or equivalently a specific unique choice of the
primordial cosmological constant, Eq.~(\ref{pcc}). This choice
must then be supplemented by a further fine-tuning -- a choice of
initial kinetic energy such that Eq.~(\ref{aes}) holds. Both
conditions are required to attain an asymptotic Einstein static
state, the one ensuring asymptotic validity of the Friedmann
equation, and the other, that of the Raychaudhuri equation.

The model can be criticized because of its fine-tuned initial
state. There are two basic responses to this criticism.
\begin{itemize} \item The criticism does not rule out these models
as valid physical models, but rather claims they are not likely to
occur in reality because they are improbable. But the inflationary
singularity theorems~\cite{bv1} are not based on probabilities,
and the Emergent models do indeed show that the geometric
conditions of those theorems need not be satisfied. \item The
force of the fine-tuning criticism is based partly in philosophy
not only in physics. There is no scientifically based proof that
the unique physical universe has to be probable. \end{itemize}

This points to a tension in cosmology between two viable but
opposed views on how to explain the current state of the universe.
\begin{itemize}
\item The one view is that the present state of the universe is
highly probable because physical processes make it very likely to
have occurred~\cite{mis}. \item The opposite view is that Nature
prefers symmetry, and the universe is likely to have originated in
a highly symmetric, and necessarily fine-tuned,
state~\cite{einstat}.  \end{itemize}

The key point is that there is no scientific proof that the one or
the other of these approaches is the correct approach to use in
cosmology. A quantum gravity theory may resolve this and other
issues relating to the origin of the universe, but in the absence
of such a theory, it seems worthwhile to pursue the implications
of both approaches.

The underlying problem within classical and semi-classical
cosmology is the uniqueness of the universe, and all the
scientific and philosophical difficulties that this
entails~\cite{unique}. One cannot straightforwardly apply
statistics or probability to a unique object. There may be a
scientific basis for the use of probability in the context of an
ensemble of universes -- a multiverse. But this is a complex and
controversial proposal, and there is no evidence that it is
correct; indeed it may not be scientifically
testable~\cite{multi}. Consequently the choice between these two
fundamentally different approaches to cosmological origins remains
of necessity a partly philosophical one, while we lack a more
complete quantum gravity theory that explains what actually
underlies the existence of the real physical universe.

Furthermore, it is not simply a case of fine-tuning versus
non-fine-tuning. The standard $K=0$ inflationary models, which
avoid a special initial state, are not free of fine-tuning, in
particular, the fine-tuning entailed in the choice of $K=0$. There
is no mechanism to attain $K=0$ (as opposed to $\Omega\to 1$), and
no obvious way to prove observationally that $\Omega =1$ to the
infinite accuracy required to establish that in reality $K=0$.
This standard assumption would appear to involve the same kind of
fine-tuning as occurs in the Emergent Universe, because of the
exact balance required to set $K=0$.

The standard inflationary models necessarily have a singularity in
the past~\cite{bv1}. Although it is possible to evade the
singularity when $K=0$~\cite{ag}, this requires complicated
modifications of the spacetime. It also appears to invoke a
fine-tuned initial state, not unlike the fine-tuning involved in
the Emergent models. The choice can therefore also be seen as
between a singularity and a fine-tuned non-singular initial state.
There are arguments for each choice. The singular inflationary
models have the strength of generality of initial conditions (at
some time after the classical singularity, i.e., effectively after
the Planck era). They have a disadvantage of starting at a
spacetime singularity where all of physics breaks down and
spacetime itself comes to an end (though quantum gravity effects
may be able to avoid this). The Emergent Universe has the
advantage of avoiding a singularity (with or without a Planck
era). There are also arguments that a highly symmetric start to
the universe is necessary for the thermodynamic arrow of time to
function as it does~\cite{penr}.

The Emergent Universe scenario gives a framework for investigating
what are the implications if whatever process caused the universe
to come into being, preferred the high-symmetry state of the
Einstein static universe to any less ordered state. There are
various arguments in support of the Einstein static model as a
preferred initial state. \begin{itemize} \item It is neutrally
stable against inhomogeneous linear perturbations when $\rho _{
\mathrm{i}}=0$ (as in the simple Emergent model above), or when
$\rho _{ \mathrm{i}}>0$ and the sound speed of matter obeys
$c_{\mathrm{s}}^{2}>{ \frac{1}{5}}$, as shown in Ref.~\cite{bemt},
extending the initial results of Refs.~\cite{hein,gein}. It is of
course unstable to \emph{homogeneous} perturbations, which break
the balance between curvature and energy density. This instability
is crucial for producing an inflationary era. \item It has no
horizon problem. \item It maximizes the entropy within the family
of FRW radiation models~\cite{gein}. \item It is the unique
highest symmetry non-empty FRW model, being invariant under a
7-dimensional group of isometries~\cite{ell67}.
\end{itemize}

The Einstein static model has a well-defined vacuum, and Casimir
effects (see, e.g., Ref.~\cite{cas}) could play an important role
in determining the primordial parameters of the Emergent Universe.

\section{Emergent models with a finite time of inflation}

The realization of the Emergent Universe outlined above and
elaborated in Ref.~\cite{emerge}, illustrated qualitatively in
Fig.~1, has the advantage of simplicity. But the fact that the
initial Einstein static state is only achieved asymptotically in
the infinite past could be seen as a disadvantage. It is possible
to find other realizations of the scenario in which the universe
starts expanding from an Einstein static state at a finite time in
the past.

For example, one can consider potentials with a critical value
$V(\phi_{\mathrm{i }})$, where $V^{\prime}(\phi_{\mathrm{i}})=0$,
above a stable minimum at $\phi_{\mathrm{f}}$, i.e.,
\begin{equation}
V(\phi_{\mathrm{f}})<V(\phi_{\mathrm{i}})\,,
~V^{\prime}(\phi_{\mathrm{f}}) = 0 <
V^{\prime\prime}(\phi_{\mathrm{f}})\,.
\end{equation}
The scalar field is initially at the critical position
$\phi=\phi_{\mathrm{i} }$; since there is no kinetic energy
($\dot{\phi}_{\mathrm{i}}=0$), matter is necessary
($\rho_{\mathrm{i}}>0$) to provide an initial static state.
Specifically, by Eqs.~(\ref{es1}) and (\ref{es2}), we have
\begin{equation}
V(\phi_{\mathrm{i}})={\frac{1}{2}}
(1+3w_{\mathrm{i}})\rho_{\mathrm{i}}\,.
\end{equation}
If the initial position is an unstable (tachyonic) maximum, i.e.,
$ V^{\prime\prime}(\phi_{\mathrm{i}})<0$, then the field rolls
down to the true minimum, similar to natural inflation potentials.
If the initial position is a false vacuum minimum, then the field
needs to tunnel towards the true minimum.

Another alternative is an inflationary potential that depends on
temperature in much the same way as in the first inflationary
universe models. In the initial Einstein static state, with matter
at some temperature, the field starts at a minimum, which becomes
unstable after a perturbation. This leads to expansion of the
universe, and consequently the temperature falls, which results in
the appearance of a lower minimum.

\section{Conclusions}

We have shown that inflationary cosmologies exist in which the
horizon problem is solved before inflation begins, there is no
singularity, no exotic physics is involved, and the quantum
gravity regime can even be avoided. These Emergent Universe models
can be constructed with simple potentials (illustrated
schematically in Fig.~1), giving past-infinite inflation from an
asymptotically initial Einstein static state, with a bounded
number of e-folds and redshift, followed by reheating in the usual
way. Explicit and simple forms of the potential can be found that
are consistent with present cosmological observations for suitable
choice of parameters in the potential~\cite{emerge}. Other
realizations exist in which the universe starts inflating from an
Einstein static state at a finite time in the past. The Emergent
models illustrate the potentially strong primordial effects of
positive spatial curvature, leading to a very different early
universe than the standard models, while producing a late universe
that can be observationally distinguished from the standard case
only by high-precision observations.

If one requires, within classical and semi-classical general
relativity theory, a non-singular universe with standard fields
and matter, then this is only possible if $K=+1$. If one further
rules out bouncing models because of problems in achieving a
regular bounce, then one is led to the Emergent Universe scenario,
with its fine-tuned initial Einstein static state. This state is
preferred by its appealing stability, entropy and geometric
properties, and provides, because of its compactness, an
interesting arena for investigating the Casimir and other effects.

The Emergent Universe scenario with $K=+1$ provides an interesting
alternative to the standard inflationary scenario with $K=0$, in
the case that $\Omega_0$ is close to, but above, 1. This may be
seen as a choice between a fine-tuned non-singular initial state,
and a singularity that precedes generic initial conditions. We do
not claim that the Emergent scenario is superior to the standard
scenario, but simply that it is worthwhile investigating, since
neither it nor the standard scenario are yet ruled out by
scientific arguments. In the absence of a quantum gravity theory
that can explain the true nature of the origin of the universe,
both approaches remain valid to investigate and test. The two
approaches lead to observationally viable models, at least until
there is firm observational evidence determining the curvature
parameter $K$ of the universe.

If future observations turn out to provide strong evidence for
positive curvature, then the standard models will be ruled out,
and the Emergent models and other closed models with a singularity
and deceleration preceding inflation~\cite{ld}, would be
contenders to describe the universe. Of course if $\Omega_0-1$ is
shown to be negative, then all closed models and the standard
models will be ruled out. However, if the Universe is in fact
exactly flat, observations will not be able to rule out a very
tiny positive curvature (and the existence of perturbations
prevents a very accurate measurement). Thus if the standard is
model is correct, the Emergent Universe with perturbatively small
curvature cannot be falsified by observations, although it does
produce a very different early universe.

~\newline

\textbf{Acknowledgements}

We thank Anthony Aguirre, John Barrow, Bruce Bassett, David Coule,
Jeff Murugan, Christos Tsagas, Tanmay Vachaspati and David Wands
for helpful comments, and Emily Leeper for assistance with the
figure. GE is supported by the NRF (SA) and RM by PPARC (UK). RM
thanks the University of Cape Town for hospitality while part of
this work was done.

\end{document}